\documentclass[a4paper,11pt]{article}

\usepackage{amsmath,amssymb,bbm,mathtools}
\usepackage{ascmac}
\usepackage{cancel}
\usepackage{xcolor} 
\usepackage[bookmarks=true,bookmarksnumbered=true,bookmarkstype=toc]{hyperref} 
\hypersetup{
pdfauthor={Tetsuji Kimura, Shin Sasaki and Masaya Yata},
colorlinks={true},
linkcolor={blue},
urlcolor={blue},
filecolor={blue},
citecolor={blue}
}
\usepackage{ulem}

\makeatletter

 \@addtoreset{equation}{section}
\makeatother

\newcounter{Enumerate}

\DeclareFontFamily{U}{rsf}{}
\DeclareFontShape{U}{rsf}{m}{n}{
  <5> <6> rsfs5 <7> <8> <9> rsfs7 <10-> rsfs10}{}
\DeclareMathAlphabet\Scr{U}{rsf}{m}{n}

\usepackage[mathscr]{eucal}

\newcommand{\del}{\partial}

\newcommand{\ls}{\ \ \ \ \ }

\newcommand{\ve}{\varepsilon}

\newcommand{\dps}{\displaystyle}
\newcommand{\bsubeq}{\begin{subequations}}
\newcommand{\esubeq}{\end{subequations}}

\newcommand{\noi}{\noindent}

\newcommand{\N}{\mathcal{N}}

\newcommand{\slb}{\scalebox}

\definecolor{royalblue}{HTML}{2B60DE}

\begin{document}
\allowdisplaybreaks{

\thispagestyle{empty}


\begin{flushright}
TIT/HEP-650 \\
\end{flushright}

\vspace{30mm}

\noi
\slb{2.5}{World-volume Effective Action of}

\vspace{5mm}

\noi
\slb{2.5}{Exotic Five-brane in M-theory 
}

\vspace{10mm}

\slb{1.2}{Tetsuji {\sc Kimura}$^{a,b}$}, \
\slb{1.2}{Shin {\sc Sasaki}$^{c}$} 
\ and \ 
\slb{1.2}{Masaya {\sc Yata}$^{d}$}

\slb{.9}{\renewcommand{\arraystretch}{1.2}
\begin{tabular}{rl}
$a$ & 
{\sl 
Research and Education Center for Natural Sciences, 
Keio University
}
\\
& {\sl 
Hiyoshi 4-1-1, Yokohama, Kanagawa 223-8521, JAPAN
}
\\
& {\tt tetsuji.kimura \_at\_ keio.jp}
\end{tabular}
}

\slb{.9}{\renewcommand{\arraystretch}{1.2}
\begin{tabular}{rl}
$b$ & 
{\sl 
Department of Physics,
Tokyo Institute of Technology} 
\\
& {\sl 
Tokyo 152-8551, JAPAN
}
\end{tabular}
}

\slb{.9}{\renewcommand{\arraystretch}{1.2}
\begin{tabular}{rl}
$c$ & {\sl
Department of Physics,
Kitasato University}
\\
& {\sl 
Sagamihara 252-0373, JAPAN}
\\
& {\tt shin-s \_at\_ kitasato-u.ac.jp}
\end{tabular}
}

\slb{.9}{\renewcommand{\arraystretch}{1.2}
\begin{tabular}{rl}
$d$ & {\sl
Department of Physics, National University of Singapore}
\\  
& {\sl 
2, Science Drive 3, Singapore 117542, Singapore}
\\
& {\tt phymasa \_at\_ nus.edu.sg}
\end{tabular}
}

\vspace{15mm}


\slb{1.1}{\sc Abstract}
\begin{center}
\slb{.95}{
\begin{minipage}{.95\textwidth}
We study the world-volume effective action of an exotic five-brane, 
known as the M-theory $5^3$-brane (M$5^3$-brane) in eleven dimensions.
The supermultiplet of the world-volume theory is the $\mathcal{N} = (2,0)$ tensor multiplet in six dimensions.
The world-volume action contains three Killing vectors 
$\hat{k}_{\hat{I}}^M \ (\hat{I} = 1,2,3)$ associated with the $U(1)^3$ isometry. 
We find the effective T-duality rule for the eleven-dimensional
 backgrounds that transforms the M5-brane effective action to that of
 the M$5^3$-brane.
We also show that our action provides the source term for the M$5^3$-brane
 geometry in eleven-dimensional supergravity.
\end{minipage}
}
\end{center}

\newpage

\section{Introduction}
It is known that there are various kinds of extended objects in string theories. 
Among other things, D-branes, the most familiar extended
objects in superstring theories, have been studied in various contexts.
There are other extended objects such as Kaluza-Klein monopoles (KK5-branes) and NS5-branes.
They are BPS solutions in supergravity theories whose background geometries are governed by
single valued functions of space-time.

A remarkable fact about these branes are that they are related by 
U-duality in string theories.
U-duality is a key property to
understand the non-perturbative definition of string theories, namely, M-theory.
BPS states in $d$-dimensional supermultiplets obtained from the M-theory 
compactified on the $(11-d)$-dimensional torus $T^{11-d}$ are interpreted as 
BPS branes wrapping cycles in the torus. 
The BPS branes consist of D-branes, NS5-branes, KK5-branes and waves.
They are standard branes which have been studied extensively.
In addition to these branes, in $d \le 8$ dimensions, 
there are other extended objects known as exotic branes \cite{Elitzur:1997zn,Obers:1998fb}.
The background geometries of the exotic branes 
are described by multi-valued functions of space-time and 
exhibit non-trivial monodromies given by the U-duality group.
This kind of geometry is known as the non-geometric background and called the U-fold \cite{Hull:2004in}.

We have studied exotic five-branes in string theories.
In particular, we have studied the exotic $5^2_2$-brane\footnote{The
$5^2_2$-brane has two isometries along transverse directions and its
tension is proportional to $g_s^{-2}$. Here $g_s$ is the string coupling
constant. See \cite{Obers:1998fb,deBoer:2010ud,deBoer:2012ma} for the notation of $5^2_2$.}
from the viewpoint of the worldsheet theories 
\cite{Kimura:2013fda, Kimura:2013zva, Kimura:2013khz, Kimura:2015yla, Kimura:2015qze}, 
the background geometries as supergravity solutions \cite{Kimura:2014wga, Kimura:2014bea},
the supersymmetry conditions \cite{Kimura:2016xzd} and the world-volume theory \cite{Kimura:2014upa}.
Remarkably, the world-volume theory is the most direct approach to study the dynamics of extended objects.
Applying the T-duality transformation to the NS5-branes twice, we
obtain the world-volume effective action of the $5^2_2$-brane in type
IIB string theory \cite{Chatzistavrakidis:2013jqa} whose dynamics is governed
by the $\mathcal{N} = (1,1)$ vector multiplet in six dimensions.
On the other hand, the type IIA $5^2_2$-brane contains the 
2-form field in the $\mathcal{N} = (2,0)$ tensor multiplet.
We derived the effective action of the type IIA $5^2_2$-brane 
\cite{Kimura:2014upa} from the type
IIA NS5-brane whose M-theory origin is the M5-brane in eleven-dimensions \cite{Pasti:1997gx}.

We know that the BPS extended objects in type IIA string theory have their 
eleven-dimensional M-theory origin.
The D0-brane is obtained from the M-wave.
The F-string and the D2-brane come from the M2-brane while the NS5-brane
and the D4-brane come from the M5-brane. 
The type IIA KK5-brane and the D6-brane are obtained by the double/direct dimensional reductions of the KK6-brane,
respectively. 
The D8-brane and also the KK8-brane is expected to come from a nine-dimensional object in
eleven dimensions \cite{Bergshoeff:1998bs, Bergshoeff:1998re}.
It is also important to study M-theory origins of exotic branes in string theories.
Indeed, there is an exotic five-brane known as the $5^3$-brane
\cite{LozanoTellechea:2000mc, deBoer:2010ud,deBoer:2012ma}
in eleven-dimensional M-theory. We call this the M$5^3$-brane in this paper.
The M$5^3$-brane has isometries along three transverse directions to its
world-volume.
The tension of the M$5^3$-brane is proportional to the volume of the
torus $T^3$ realized in the isometry directions.
This is just a generalized KK monopole in eleven dimensions \cite{Bergshoeff:1997gy}. 
The supergravity solution associated with the M$5^3$-brane exhibits non-geometric nature whose
monodromy is given by the U-duality group $SL(3,\mathbb{Z}) \times
SL(2,\mathbb{Z})$ in eight dimensions.
The direct/double dimensional reductions of the M$5^3$-brane to ten dimensions
provide the exotic $5^2_2$-brane/$4^3_3$-brane in type IIA string theory.
Therefore the M$5^3$-brane is the higher dimensional origin of the type IIA exotic
branes. In this paper we study the world-volume effective action of the exotic
M$5^3$-brane in eleven dimensions.

The organization of this paper is as follows.
In the next section, we introduce the world-volume effective action of
the exotic $5^2_2$-brane in type IIA string theory.
In section 3, we propose the world-volume effective action of the M$5^3$-brane.
We will show that the action correctly reproduces the action 
of the type IIA $5^2_2$-brane in ten dimensions.
In section 4, we will show that the actions of the $5^2_2$-brane and
M$5^3$-brane give the source terms of these five-brane geometries.
Section 5 is devoted to the conclusion and discussions.

\section{Effective action of type IIA $5^2_2$-brane}
In this section we introduce the world-volume effective action of the
exotic $5^2_2$-brane in type IIA string theory.
The background geometry of the $5^2_2$-brane is a 
1/2-BPS solution to the type IIA supergravity and
the world-volume theory has 16 conserved supercharges.
The geometry of the $5^2_2$-brane exhibits the monodromy given by the 
$SO(2,2)$ T-duality group and this is a T-fold.

The world-volume action of the $5^2_2$-brane is obtained from that of the type IIA NS5-brane 
by repeated T-duality transformations.
The type IIA NS5-brane is obtained by the direct dimensional reduction of the M5-brane in eleven dimensions. 
We start from the effective action of the M5-brane which has been established in \cite{Pasti:1997gx}.
The supermultiplet of the six-dimensional world-volume theory on 
the M5-brane is the $\mathcal{N} = (2,0)$ tensor multiplet.
This consists of a 2-form field $A_{ab}$, five scalar
fields $X^{\hat{M}} \ (\hat{M} = 1,\ldots, 5)$ and their
superpartners. The field strength 
$F_{abc} = \partial_a A_{bc} - \partial_b A_{ac} + \partial_c
 A_{ab} \ (a,b,c = 0, \ldots,5)$ satisfies the self-duality condition in six dimensions.
The effective action of the M5-brane\footnote{We always consider the actions for the bosonic fields in this paper.}, the so-called Pasti-Sorokin-Tonin (PST) action, is
\begin{align}
S_{\text{M5}} =& \ 
- T_{\text{M5}} \int \! d^6 \xi \ 
\left[
\sqrt{
- \det (P[\hat{g}]_{ab} + i \hat{H}^{*}_{ab})
}
+ \frac{\sqrt{- \hat{g}}}{4} (\partial_a a \partial^a a)^{-1}
 \hat{H}^{*abc} \hat{H}_{bcd} \partial_a a \partial^d a
\right] + S_{\mathrm{WZ}},
\label{eq:M5_action}
\end{align}
where $T_{\mathrm{M5}} = \frac{1}{(2\pi)^5} M_{11}^6 $ is the tension of the M5-brane and 
$M_{11}$ is the Planck mass in eleven dimensions.
The symbol $P$ stands for the pull-back of the background fields:
\begin{align}
P[\hat{g}]_{ab} = \hat{g}_{MN} \partial_a X^M \partial_b X^N, \qquad 
P[\hat{C}^{(3)}]_{abc} = \hat{C}^{(3)}_{MNP} \partial_a X^M \partial_b
 X^N \partial_c X^P,
\end{align}
where $\hat{g}_{MN} \ (M,N = 0,1,\ldots,10)$ and $\hat{C}^{(3)}$ are the metric and the 3-form
in eleven-dimensional supergravity.
$X^M$ and $\xi^a \ (a=0,\ldots,5)$ are the space-time and world-volume
coordinates, respectively.
$\hat{g}_{ab} = P[\hat{g}]_{ab}$ is the induced metric on the
M5-brane world-volume and $\hat{g}$ is the determinant of
$\hat{g}_{ab}$.
The world-volume indices $a,b,c, \ldots = 0, \ldots, 5$ are raised and lowered 
by the induced metric $\hat{g}_{ab}$ and its inverse, $\hat{g}^{ab}$.
We note that the action \eqref{eq:M5_action} contains a non-dynamical auxiliary field
$a$ which is needed to write down the action for the self-dual field in a Lorentz covariant way.
The world-volume 2-form field $A_{ab}$ enters into the action
\eqref{eq:M5_action} with the following combinations:
\begin{align}
& \hat{H}_{abc} = F_{abc} - P[\hat{C}^{(3)}]_{abc}, \notag \\
& \hat{H}^{*abc} = \frac{1}{3!} \frac{1}{\sqrt{- \hat{g}}}
 \varepsilon^{abcdef} \hat{H}_{def}, \notag \\
& \hat{H}^{*}_{ab} = \frac{1}{\sqrt{
\partial_g a \partial^g a}}
 \hat{H}^{*}_{abc} \partial^c a 
= \frac{1}{3!} \frac{1}{\sqrt{- \hat{g}}} \frac{1}{\sqrt{
\partial_g a \partial^g a}} \varepsilon_{ab} {}^{cdef} \hat{H}_{def} \partial_c a.
\end{align}
Here $\varepsilon_{abcdef}$ is the Levi-Civita symbol.
The five scalars $X^{\hat{M}}$, which appear in the pull-back of the
background fields in the static gauge, represent the transverse
fluctuation modes (named the geometric zero-modes) of the M5-brane in eleven dimensions.
The M5-brane action \eqref{eq:M5_action} has symmetries of the
world-volume $U(1)$ gauge transformations, two field dependent
world-volume gauge transformations and the space-time gauge
transformations of the background fields \cite{Pasti:1997gx, Isono:2014bsa, Bandos:2014bva}.
In the following, we do not consider the Wess-Zumino part $S_{\mathrm{WZ}}$.

We now perform the direct dimensional reduction of the M5-brane action
\eqref{eq:M5_action} to ten dimensions.
We decompose the eleven-dimensional index $M = (\mu, \sharp)$ where $\mu
= 0, \ldots, 9$ and $\sharp = 10$. 
The former represents the ten-dimensional space-time, and the latter
is the compactified M-circle direction.
The KK ansatz of the eleven-dimensional metric $\hat{g}_{MN}$ and the
3-form $\hat{C}^{(3)}_{MNP}$ is 
\begin{align}
\hat{g}_{MN} = 
\left(
{\renewcommand{\arraystretch}{1.25}
\begin{array}{cc}
e^{- \frac{2}{3} \phi} (g_{\mu \nu} + e^{2\phi} C^{(1)}_{\mu}
 C^{(1)}_{\nu}) & e^{\frac{4}{3} \phi} C^{(1)}_{\mu} \\
e^{\frac{4}{3} \phi} C^{(1)}_{\nu} & e^{\frac{4}{3} \phi}
\end{array}
}
\right), \quad \hat{C}^{(3)}_{\mu \nu \rho} = C^{(3)}_{\mu \nu \rho}, \quad
 \hat{C}^{(3)}_{\mu \nu \sharp} = - B_{\mu \nu}.
\label{eq:KKansatz}
\end{align}
Here $g_{\mu \nu}$ is the ten-dimensional metric, $\phi$ is the dilaton,
$C^{(1)}_{\mu}, C^{(3)}_{\mu \nu \rho}$ are the R-R 1- and 3-forms.
Applying this dimensional reduction to \eqref{eq:M5_action}, 
we obtain the action of the type IIA NS5-brane \cite{Bandos:2000az, Kimura:2014upa}.
The NS5-brane action has symmetries that are inherited from those of the
M5-brane. The world-volume theory of the NS5-brane consists of the four
 scalar fields $X^{I'} \ (I'=1,\ldots,4)$ which are the geometric zero-modes,
an extra scalar field $Y$ which originates from the geometric zero-mode along the M-circle direction, a 2-form 
field whose field strength satisfies the self-duality condition in six
dimensions and an auxiliary field. 
They comprises the six-dimensional $\mathcal{N} = (2,0)$ tensor multiplet as expected.

Now we introduce the Killing vector $k_1^{\mu}$ associated with the
$U(1)$ isometry of the background fields $g_{\mu \nu}, B_{\mu \nu},
\phi$, $C^{(1)}, C^{(3)}$.
This $U(1)$ isometry is defined along the transverse direction to the
NS5-brane world-volume.
Applying the T-duality transformation to the type IIA NS5-brane action 
 along this direction, we obtain the effective action of the type IIB
 KK5-brane \cite{Eyras:1998hn}.
In addition to the couplings to the background fields in type IIB
supergravity, the effective action of the KK5-brane contains couplings to the Killing
vector $k_1^{\mu}$. 
Geometrically, this corresponds to the isometry in the Taub-NUT space
and there is no geometric zero-mode associated with this direction.
The world-volume theory has three geometric zero-modes together with a 
scalar mode $Y$.
Since the world-volume theory of the type IIB KK5-brane is governed by
the $\mathcal{N} = (2,0)$ tensor multiplet, one needs one extra scalar field $\varphi_1$.
This naturally appears in the T-duality transformation as the dual coordinate of 
the isometry direction. This is nothing but the Lagrange multiplier in the dualized
process in the string world-sheet theory (see Appendix in \cite{Kimura:2014upa}).
We call $\varphi_1$ the winding zero-mode.

We next introduce the second Killing vector $k^{\mu}_2$ associated with the
$U(1)$ isometry along another transverse direction to the KK5-brane.
By further T-duality transformation, 
the KK5-brane becomes the type IIA exotic $5^2_2$-brane.
The effective action of the type IIA $5^2_2$-brane was derived in
\cite{Kimura:2014upa}. This is given by 
\begin{align}
& S^{\text{IIA}}_{5^2_2} 
= - T_{5} \int \! d^6 \xi \ 
(\det h^{[2]}) e^{-2\phi} 
\sqrt{
- \det 
(
\check{g}_{ab} + \lambda^2 (\det h^{[2]})^{-1} e^{2\phi}
 \check{F}^{(1)}_a \check{F}^{(1)}_b
)
} 
\notag \\
& \times 
\sqrt{
\det 
\Big(
\delta_a {}^b + \frac{i e^{\phi}}{3! \check{\mathcal{N}} \sqrt{(\det
 h^{[2]}) (\check{g}^{cd} \partial_c a \partial_d a)}} Z_a {}^b
\Big)
}
\notag \\
& - \frac{\lambda^2}{4} T_{5} 
\int \! d^6 \xi \ \frac{1}{3! \check{\mathcal{N}}^2} 
\frac{1}{\check{g}^{pq} \partial_p a \partial_q a} 
\varepsilon^{abcdef} \check{H}_{def} \check{H}_{abg} \partial_c a
 \partial_h a 
\left[
\check{g}^{gh} - \frac{\lambda^2 e^{2\phi} \check{g}^{gr} \check{g}^{hs}
 \check{F}^{(1)}_r \check{F}^{(1)}_s}{(\det h^{[2]}) + \lambda^2 e^{2\phi}
 \check{g}^{uv} \check{F}^{(1)}_u \check{F}^{(1)}_v}
\right],
\label{eq:IIA522}
\end{align}
where $\lambda = 2 \pi \alpha'$ is the string slope parameter, 
and $T_{5} = \frac{1}{(2\pi)^5} g_s^{-2} \alpha^{\prime -3}$ 
is the tension of the NS5-brane.
Here $\check{g}_{\mu \nu}$ is the ten-dimensional metric on which 
the T-duality transformations with the two Killing vectors $k_1^{\mu},
k_2^{\mu}$ have been applied.
The explicit form of $\check{g}_{\mu \nu}$ is 
\begin{align}
\check{g}_{\mu \nu} = & \ g_{\mu \nu} - \frac{1}{2} h^{[2]IJ} 
\left[
\left\{
i_{k_I} g - ( i_{k_I} B - \lambda d \varphi_I)
\right\}_{\mu}
\left\{
i_{k_J} g + (i_{k_J} B - \lambda d \varphi_J)
\right\}_{\nu}
+ (\mu \leftrightarrow \nu)
\right],
\label{eq:gTdual_final}
\end{align}
where we have defined the ``Killing matrix'' 
$h^{[2]}_{IJ} = (g_{\mu \nu} + B_{\mu \nu}) k_I^{\mu} k_J^{\nu}$ 
and its inverse $h^{[2]IJ}$ (where $I,J = 1,2$).
The inner product of an $n$-form $X^{(n)}$ with a Killing vector
$k^{\mu}$ is defined by 
\begin{align}
(i_{k} X^{(n)})_{\mu_1 \cdots \mu_{n-1}} = k^{\nu} X^{(n)}_{\nu
 \mu_1 \cdots \mu_{n-1}}.
\end{align}
The effective induced metric is defined by $\check{g}_{ab} =
P[\check{g}]_{ab}$, and $\check{g}^{ab}$ is the inverse of
$\check{g}_{ab}$.
Note that in the pull-back, we define $P[d \varphi_I]_a = \partial_a \varphi_I$.
In the action (\ref{eq:IIA522}), we have defined the following quantities:
\begin{align}
Z_a {}^b =& \ 
\frac{
\varepsilon^{gbcdef} 
(
\check{g}_{ac} + \lambda^2 (\det h^{[2]})^{-1} e^{2\phi}
 \check{F}^{(1)}_a \check{F}^{(1)}_c 
)
\check{H}_{def} \partial_g a
}
{
- \det 
(
\check{g}_{pq} + \lambda^2 (\det h^{[2]})^{-1} e^{2\phi}
 \check{F}^{(1)}_p \check{F}^{(1)}_q
)
}, 
\notag \\
\check{F}^{(1)}_a =& \ \partial_a Y + \lambda^{-1}
 P[\check{C}^{(1)}]_a, 
\notag \\
\check{\mathcal{N}}^2 =& \ 
1 - \frac{
\lambda^2 e^{2\phi} (\check{g}^{ef} \check{F}^{(1)}_e \partial_f a )^2
}
{
(\check{g}^{ab} \partial_a a \partial_b a)
(
(\det h^{[2]}) + \lambda^2 e^{2\phi} \check{g}^{cd} \check{F}^{(1)}_c \check{F}^{(1)}_d
)
}, 
\notag \\
\check{H}_{abc} =& \ F_{abc} - P[\check{C}^{(3)}]_{abc} - \lambda (P[\check{B}]
 \wedge \check{F}^{(1)})_{abc}.
\label{eq:522quantities}
\end{align}
Here $\check{B}, \check{C}^{(1)}, \check{C}^{(3)}$ are the NS-NS B-field, R-R
1- and 3-forms on which the T-duality transformations have been applied.
Their explicit forms are
\begin{align}
\check{B} =& \ B - \frac{1}{2} h^{[2]IJ} 
\left\{
i_{k_I} g - (i_{k_I} B - \lambda d \varphi_I)
\right\}
\wedge 
\left\{
i_{k_J} g + (i_{k_J} B - \lambda d \varphi_J)
\right\}, 
\label{eq:BTdual_final}
\notag \\
\check{C}^{(1)} =& \ - i_{k_1} i_{k_2} C^{(3)} - (i_{k_1} i_{k_2} B)
 C^{(1)} + \epsilon^{IJ} (i_{k_I} C^{(1)}) (i_{k_J} B - \lambda d
 \varphi_J), \\
\check{C}^{(3)} =& \ - i_{k_1} i_{k_2} C^{(5)} - (i_{k_1} i_{k_2} B)
 C^{(3)} + \epsilon^{IJ} (i_{k_I} C^{(3)}) \wedge (i_{k_J} B - \lambda d
 \varphi_J)
\notag \\
& \ - \frac{1}{2} \epsilon^{IJ} C^{(1)} \wedge (i_{k_I} B - \lambda d \varphi_I) \wedge
 (i_{k_J} B - \lambda d \varphi_J)
\notag \\
& \ - \frac{1}{2}
\left[
\frac{}{}
i_{k_1} i_{k_2} C^{(3)} + (i_{k_1} i_{k_2} B) C^{(1)} 
- \epsilon^{KL} (i_{k_K} C^{(1)}) (i_{k_L} B - \lambda d \varphi_L)
\right] 
\notag \\
& \qquad \qquad \wedge h^{[2]IJ} 
\left\{
i_{k_I} g - (i_{k_I} B - \lambda d \varphi_I)
\right\} 
\wedge 
\left\{
i_{k_J} g + (i_{k_J} B - \lambda d \varphi_J)
\right\}.
\end{align}
Here we introduced an antisymmetric symbol $\epsilon^{IJ} = - \epsilon^{JI}$.
Since the $5^2_2$-brane is a defect brane of co-dimension two \cite{Bergshoeff:2011se}, 
there are only two geometric zero-modes in the world-volume action.
Indeed, the effective theory contains two scalar fields $X^1, X^2$ in
the pull-back of the 
type IIA supergravity backgrounds \cite{Kimura:2014upa}.
There are also a scalar field $Y$ which comes from
the M-circle and a 2-form field $A_{ab}$.
In addition, there are two winding zero-modes $\varphi_1, \varphi_2$ introduced in the T-duality process.
Since the factor $\det h^{[2]}$ contains the volume of the 2-torus $T^2$,
the tension of the $5^2_2$-brane is proportional to 
$g_s^{-2} (\text{vol} (T^2))$ which is consistent with the definition of
the $5^2_2$-brane.
In our previous paper \cite{Kimura:2014upa}, we have discussed the gauge
symmetries of the action \eqref{eq:IIA522} and its Wess-Zumino coupling.

The action \eqref{eq:IIA522} should be obtained from the direct
dimensional reduction of the action of the M$5^3$-brane. 
In the next section, we work out the M$5^3$-brane action which is
consistent with the action \eqref{eq:IIA522}.

\section{Effective action of M$5^3$-brane}
In this section, we study the M$5^3$-brane effective action in the
eleven-dimensional supergravity background.
We concentrate on the purely geometric background where the 3-form field $\hat{C}^{(3)}$ is absent.
The effective theory has the following properties.
(i) The world-volume theory is governed by the $\mathcal{N} = (2,0)$
tensor multiplet in six dimensions. The bosonic fields are 
a 2-form $A_{ab}$ whose field strength satisfies the self-duality
condition, two scalar fields $X^1, X^2$ that correspond to the geometric zero-modes,
three scalar fields $\hat{\varphi}^{\hat{I}}$ $(\hat{I} = 1,2,3)$ that
correspond to the dual winding coordinates and an auxiliary field $a$.
(ii) The background is the eleven-dimensional metric $\hat{g}_{MN}$ with
the $U(1)^3$ isometry.
(iii) The M$5^3$-brane couples to three Killing vectors
$\hat{k}^M_{\hat{I}}$ associated with the $U(1)^3$ isometry. 
The effective tension of the M$5^3$-brane is proportional to the volume
of $T^3$ where the $U(1)^3$ isometry is realized.
(iv) After the direct dimensional reduction to ten dimensions, the
action should reproduce the $5^2_2$-brane action 
where the NS-NS B-field and the R-R 3-form $C^{(3)}$ are turned off.

In the following, we explore the structure of the M$5^3$-brane action in
each world-volume field sector. 
In particular, we verify that the proposed structure is consistent with the type IIA $5^2_2$-brane action.

\subsubsection*{Geometric zero-mode sector}
We begin with the following terms in the first line of the type IIA $5^2_2$-brane effective
action \eqref{eq:IIA522}:
\begin{align}
\check{g}_{ab} + \lambda^2 (\det h^{[2]})^{-1} e^{2\phi}
 \check{F}^{(1)}_a \check{F}^{(1)}_b
\, .
\label{eq:IIA_DBI}
\end{align}
When $C^{(3)} = B = 0$, \eqref{eq:IIA_DBI} becomes 
\begin{align}
& \ \Pi^{[2]}_{\mu \nu} (k_1,k_2) \partial_a X^{\mu} \partial_b X^{\nu} +  
\frac{\lambda^2}{\det h^{[2]}} g_{\mu \nu} (k_1^{\mu} \partial_a \varphi_2 -
 k_2^{\mu} \partial_a \varphi_1)
(
k_1^{\nu} \partial_b \varphi_2 - k_2^{\nu} \partial_b \varphi_1
)
\notag \\
& \ + 
\frac{\lambda^2 e^{2\phi}}{\det h^{[2]}}
\left[ \frac{}{} \right.
\partial_a Y \partial_b Y 
- 
\left\{
(i_{k_1} C^{(1)}) \partial_a \varphi_2 - (i_{k_2} C^{(1)}) \partial_a \varphi_1
\right\} \partial_b Y 
- 
\left\{
(i_{k_1} C^{(1)}) \partial_b \varphi_2 - (i_{k_2} C^{(1)}) \partial_b \varphi_1
\right\} \partial_a Y 
\notag \\
& \
+ 
\left\{
(i_{k_1} C^{(1)}) \partial_a \varphi_2 - (i_{k_2} C^{(1)}) \partial_a \varphi_1
\right\}
\left\{
(i_{k_1} C^{(1)}) \partial_b \varphi_2 - (i_{k_2} C^{(1)}) \partial_b \varphi_1
\right\}
\left. \frac{}{} \right],
\label{eq:extra_scalars}
\end{align}
where $\Pi^{[2]}_{\mu \nu} (k_1, k_2)$ is the projector constructed by $k_1^{\mu}, k_2^{\mu}$:
\begin{align}
\Pi^{[2]}_{\mu \nu} (k_1, k_2) = 
g_{\mu \nu} 
- h^{[2]IJ} \, k_I^{\rho} k_J^{\sigma} \, g_{\mu \rho} g_{\nu \sigma}.
\end{align}
Here $h^{[2]IJ}$ is the inverse of 
$h^{[2]}_{IJ} = k^{\mu}_I k^{\nu}_J g_{\mu \nu}$.
We first look for the M$5^3$-brane origin of the first term in
\eqref{eq:extra_scalars}, i.e., the geometric zero-mode sector.
Since the M$5^3$-brane is a defect brane and has the $U(1)^3$ isometry in the transverse
directions, there are no geometric zero-modes in these directions.
Only two geometric zero-modes which correspond to the two transverse
directions appear in the pull-back.
We expect that the two scalar fields associated with the geometric
zero-modes are described by the gauged sigma model \cite{Bergshoeff:1997gy}.
We start from the Nambu-Goto action as the building
block of the M$5^3$-brane world-volume action:
\begin{align}
S = - T_{\text{M}5} \int \! d^6 \xi \ \sqrt{
- \det (\hat{g}_{MN} \partial_a X^M \partial_b X^N)
}.
\label{eq:Nambu-Goto}
\end{align}
In order to obtain the gauged sigma model, we 
introduce an auxiliary world-volume gauge fields $C^{\hat{I}}_a \ (\hat{I}=1,2,3)$ and gauge
covariantize the pull-back in the action \eqref{eq:Nambu-Goto}:
\begin{align}
D_a X^M = \partial_a X^M + C^{\hat{I}}_a
 \hat{k}^{M}_{\hat{I}}.
\end{align}
With this covariantization, the isometry of the background metric
$\hat{g}_{MN}$ becomes a gauge symmetry in the world-volume.
After integrating out the auxiliary gauge field $C_a^{\hat{I}}$, 
the metric $\hat{g}_{MN}$ in \eqref{eq:Nambu-Goto} becomes
\begin{align}
\hat{\Pi}^{[3]}_{MN} (k) =& \  
\hat{g}_{MN} 
- \hat{h}^{[3]\hat{I} \hat{J}} \, \hat{k}_{\hat{I}}^P
 \hat{k}_{\hat{J}}^Q \, \hat{g}_{PM} \hat{g}_{QN}.
\end{align}
This is a projector which projects out the three geometric 
zero-modes associated with the isometries.
Here we have defined the $3 \times 3$
Killing matrix $\hat{h}^{[3]}_{\hat{I} \hat{J}}$ and its inverse
$\hat{h}^{[3] \hat{I} \hat{J}}$ in eleven dimensions:
\begin{align}
\hat{h}^{[3]}_{\hat{I} \hat{J}} = \hat{g}_{MN} \hat{k}_{\hat{I}}^M
 \hat{k}_{\hat{J}}^N, \qquad 
\hat{h}^{[3]\hat{I} \hat{J}} \hat{h}^{[3]}_{\hat{J} \hat{K}} = 
 \delta^{\hat{I}} {}_{\hat{K}}.
\end{align}
The action \eqref{eq:0th} contains two geometric zero-modes.
As a generalization of the effective action of the KK6-brane
\cite{Bergshoeff:1997gy}, the action of the M$5^3$-brane has the overall factor 
$(\det \hat{h}^{[3]}_{\hat{I} \hat{J}})$ which guarantees the correct tension.
Indeed, this factor gives the volume of the torus $T^3$ when an
appropriate coordinate system for $\hat{k}^M_{\hat{I}}$ is employed.
Then the would-be action becomes
\begin{align}
S = - T_{\text{M}5} \int \! d^6 \xi \ 
(\det \hat{h}^{[3]})
\sqrt{
- \det \Big(
\hat{\Pi}^{[3]}_{MN} (k) \partial_a X^M \partial_b X^{N}
\Big)
}.
\label{eq:1st}
\end{align}
In order to confirm plausibility of the action \eqref{eq:1st}, 
we perform the direct dimensional reduction of \eqref{eq:1st} to ten dimensions.
We decompose the index $M = (\mu,\sharp)$ and define the M-circle
direction as $X^{\sharp} = Y$ and the corresponding Killing vector
$\hat{k}_3^M = \delta^M_{\sharp}$.
The other Killing vectors $\hat{k}_{1,2}^{M}$ become those in the
ten-dimensional $k_{1,2}^{\mu}$:
\begin{align}
\hat{k}_I^{\mu} = k^{\mu}_I, \quad \hat{k}_I^{\sharp} = 0, \quad (I,J=1,2).
\label{eq:Killing_reduction}
\end{align}
Then the dimensional reduction of the Killing matrix is given by 
\begin{align}
\hat{h}^{[3]}_{\hat{I} \hat{J}} = & \ 
\left(
{\renewcommand{\arraystretch}{1.25}
\begin{array}{cc}
e^{- \frac{2}{3} \phi} (g_{\mu \nu} + e^{2\phi} C^{(1)}_{\mu}
 C^{(1)}_{\nu}) k^{\mu}_I k^{\nu}_J & e^{\frac{4}{3} \phi} C^{(1)}_{\mu}
 k^{\mu}_I \\
e^{\frac{4}{3} \phi} C^{(1)}_{\mu} k^{\mu}_J & e^{\frac{4}{3} \phi}
\end{array}
}
\right).
\label{eq:hath_reduction}
\end{align}
We can show that the dimensional reduction of the overall volume factor of
$T^3$ is evaluated as  
\begin{align}
\det \hat{h}^{[3]} = \det h^{[2]}.
\label{eq:deth}
\end{align}
We note that the R-R 1-form $C^{(1)}$ and the dilaton dependence 
cancel out in the dimensional reduction of $\det \hat{h}^{[3]}$.

We next calculate the dimensional reduction of the projector $\hat{\Pi}^{[3]}_{MN}$.
Using the fact that $\hat{\Pi}^{[3]}_{MN}$ is the projector along
$X^{\sharp} = Y$ direction and the reduction rule for the inverse of the matrix
\eqref{eq:hath_reduction}, we find 
\begin{align}
\hat{\Pi}^{[3]}_{MN} (k) \partial_a X^M \partial_b X^N = 
e^{- \frac{2}{3} \phi} \Pi^{[2]}_{\mu \nu} (k_1,k_2) \partial_a X^{\mu}
 \partial_b X^{\nu}
\, .
\end{align}
Again, except for the overall dilaton factor $e^{-\frac{2}{3} \phi}$, 
the R-R 1-form $C^{(1)}$ and the dilaton dependence cancel out in
the reduction. Then we find that the direct dimensional reduction of \eqref{eq:1st} is given
by 
\begin{align}
-T_{5} \int \! d^6 \xi 
e^{-2\phi} (\det h^{[2]}) 
\sqrt{
- \det \Big(
\Pi^{[2]}_{\mu \nu} (k_1,k_2) \partial_a X^{\mu} \partial_b X^{\nu} 
\Big)
},
\label{eq:geometric_zeromodes}
\end{align}
where we have used the relation $T_{5} = T_{\text{M5}}$.
The action \eqref{eq:geometric_zeromodes} is nothing but the geometric
zero-modes sector of the first line in \eqref{eq:IIA522} where the
B-field and the R-R 3-form $C^{(3)}$ are turned off.
We note that the correct volume factor of the torus $T^2$ and the dilaton factor
$e^{-2\phi}$ emerges in the action.

\subsubsection*{Winding zero-modes sector}
We next consider the winding zero-mode sector.
Since $\hat{\Pi}^{[3]}_{MN} (k) \partial_a X^M \partial_b X^N$ contains only two
scalar fields, we need three extra scalar fields
$\hat{\varphi}_{\hat{I}} \ (\hat{I} = 1,2,3)$ for the $\mathcal{N} =
(2,0)$ tensor multiplet. They are interpreted as winding zero-modes 
of membranes wrapped on the three torus $T^3$ \cite{Sen:1995cf,Duff:2015jka}.
In order to see how these scalar fields appear in
the M$5^3$-brane action, we examine the terms that contain two winding
zero-modes $\varphi_1, \varphi_2$ and the M-circle scalar $Y$ in \eqref{eq:extra_scalars}.
Since the ten-dimensional metric $g_{\mu \nu}$ and the R-R 1-form
$C^{(1)}$ come from the eleven-dimensional metric $\hat{g}_{MN}$, 
we expect that the higher dimensional origin of the 
terms that contain $(\varphi_1, \varphi_2,Y)$ in
\eqref{eq:extra_scalars} is given by
\begin{align}
I_{ab} = \frac{1}{\det \hat{h}^{[3]}} \hat{g}_{MN} (\hat{k}_{\hat{I}}^M
 \partial_a \hat{\varphi}^{\hat{I}}) (\hat{k}_{\hat{J}}^N \partial_b
 \hat{\varphi}^{\hat{J}}).
\label{eq:winding_modes}
\end{align}
In order to confirm this proposal, we perform the direct dimensional
reduction of the term \eqref{eq:winding_modes}.
Using the KK ansatz \eqref{eq:KKansatz} of the metric, we find 
\begin{align}
I_{ab} =& \  
\ \frac{\lambda^2}{\det h^{[2]}} e^{- \frac{2}{3} \phi} g_{\mu \nu}
 (k_1^{\mu} \partial_a \varphi_2 - k_2^{\mu} \partial_a \varphi_1) (k_1^{\nu}
 \partial_b \varphi_2 - k_2^{\nu} \partial_b \varphi_1)
\notag \\
& \ + \frac{\lambda^2}{\det h^{[2]}} e^{\frac{4}{3} \phi} 
\left\{
(i_{k_1} C^{(1)}) \partial_a \varphi_2 - (i_{k_2} C^{(1)}) \partial_a \varphi_1
\right\}
\left\{
(i_{k_1} C^{(1)}) \partial_b \varphi_2 - (i_{k_2} C^{(1)}) \partial_b \varphi_1
\right\}
\notag \\
& \ - \frac{\lambda^2}{\det h^{[2]}
} e^{\frac{4}{3} \phi} 
\left\{
(i_{k_1} C^{(1)}) \partial_a \varphi_2 - (i_{k_2} C^{(1)}) \partial_a \varphi_1
\right\} \partial_b Y 
\notag \\
& \ - \frac{\lambda^2}{\det h^{[2]}} e^{\frac{4}{3} \phi} 
\left\{
(i_{k_1} C^{(1)}) \partial_b \varphi_2 - (i_{k_2} C^{(1)}) \partial_b \varphi_1
\right\} \partial_a Y
\notag \\
& \ 
+ \frac{\lambda^2}{\det h
^{[2]}} e^{\frac{4}{3} \phi} \partial_a Y \partial_b Y,
\end{align}
where we have identified the three scalars and the winding zero-modes as 
$\hat{\varphi}^{\hat{I}} = \lambda (\varphi_2, - \varphi_1, - Y)$.
This precisely reproduces the $(\varphi_1, \varphi_2,Y)$ terms in 
\eqref{eq:extra_scalars} except for the overall dilaton factor.
In the determinant of the $6 \times 6$ matrix $I_{ab}$, we 
extract the dilaton factor $e^{- \frac{2}{3} \phi}$ and leave it outside
of the square root.
Then the overall factor of the action becomes $e^{-2\phi}$ which gives
the correct tension of the $5^2_2$-brane.
Therefore we have confirmed that the $(\varphi_1,\varphi_2,Y)$ terms in
\eqref{eq:extra_scalars} emerge from \eqref{eq:winding_modes}.
Then the geometric and the winding zero-mode sectors of the M$5^3$-brane action are given by 
\begin{align}
S = - T_{\text{M}5} \int \! d^6 \xi \ (\det \hat{h}^{[3]}) 
\sqrt{
- \det \Big(
\hat{\Pi}^{[3]}_{MN} (k) \partial_a X^M \partial_b X^N + 
\frac{1}{\det \hat{h}^{[3]}} \hat{g}_{MN} (\hat{k}^M_{\hat{I}} \partial_a
 \hat{\varphi}^{\hat{I}}) (\hat{k}^N_{\hat{J}} \partial_b
 \hat{\varphi}^{\hat{J}})
\Big)
}.
\end{align}

\subsubsection*{Gauge and auxiliary fields sector}
Next, we study the gauge field sector in the M$5^3$-brane action.
This is the eleven-dimensional origin of the second line in \eqref{eq:IIA522}.
In the type IIA $5^2_2$-brane action, 
the 2-form gauge field $A_{ab}$ contributes to the following terms
\begin{align}
\delta^a {}_b + \frac{i e^{\phi}}{3! \check{\mathcal{N}} \sqrt{(\det h^{[2]})
 (\check{g}^{cd} \partial_c a \partial_d a )}}
Z_a {}^b,
\label{eq:Z_factor}
\end{align}
where $Z_a {}^b$ and $\check{\mathcal{N}}$ are defined in \eqref{eq:522quantities}.
We first look for the higher dimensional origin of $Z_a {}^b$.
From the discussion in the geometric and the winding zero-mode sectors,
we know that the following quantity in eleven-dimensions
\begin{align}
\hat{G}_{ab} =
\hat{\Pi}^{[3]}_{MN} (k) \partial_a X^M \partial_b X^N + 
\frac{1}{\det \hat{h}^{[3]}} \hat{g}_{MN} (\hat{k}^M_{\hat{I}} \partial_a
 \hat{\varphi}^{\hat{I}}) (\hat{k}^N_{\hat{J}} \partial_b
 \hat{\varphi}^{\hat{J}})
\label{eq:11d_metric}
\end{align}
is dimensionally reduced to 
\begin{align}
\hat{G}_{ab} =& \ 
e^{- \frac{2}{3} \phi}
\left(
\check{g}_{ab} + \lambda^2 (\det h^{[2]})^{-1} e^{2\phi} \check{F}^{(1)}_a
 \check{F}^{(1)}_b
\right).
\end{align}
We therefore expect that the M$5^3$-brane counterpart of $Z_a {}^b$ is
given by 
\begin{align}
\hat{Z}_a {}^b = \frac{\varepsilon^{gbcdef} \hat{G}_{ac} \hat{\mathcal{H}}_{def}
 \partial_g a}{\sqrt{ - \det \hat{G}}}.
\end{align}
Here $\hat{\mathcal{H}}_{abc}$ should be dimensionally reduced to $\check{H}_{abc}$
in the type IIA $5^2_2$-brane action.
We find this is given by
\begin{align}
\hat{\mathcal{H}}_{abc} =& \ F_{abc} + 3! \mathcal{G}_{3[a} \partial_b \hat{\varphi}^1
 \partial_{c]} \hat{\varphi}^2 
+ 3! \sum_{I=1,2} \hat{h}^{[3]1I} \mathcal{G}_{I[a} \partial_b \hat{\varphi}^2
 \partial_{c]} \hat{\varphi}^3
- 3! \sum_{I=1,2} \hat{h}^{[3]2I} \mathcal{G}_{I[a} \partial_b \hat{\varphi}^3
 \partial_{c]} \hat{\varphi}^1,
\label{eq:H_T-dual}
\end{align}
where we have defined
\begin{align}
\mathcal{G}_{Ia} =& \ (i_{k_I} \hat{g})_{a} - 
\frac{\hat{h}^{[3]}_{I3}}{\hat{k}_3^2}
(i_{k_3} \hat{g})_a, \qquad (I=1,2),
\notag \\
\mathcal{G}_{3a} =& \ \partial_a \hat{\varphi}^3 + \frac{1}{\hat{k}^2_3} (i_{k_3} \hat{g})_a.
\end{align}
Here $\hat{k}_3^2 = \hat{g}_{MN} \hat{k}^M_3 \hat{k}^N_3$.
Now, it is straightforward to confirm that $\hat{Z}_a {}^b$ is dimensionally
reduced to $Z_a {}^b$ in the following way,
\begin{align}
\hat{Z}_a {}^b = e^{\frac{4}{3} \phi} Z_a {}^b.
\label{eq:hat_Z_reduction}
\end{align}
We next examine the factor $\check{\mathcal{N}} \sqrt{(\det h^{[2]})
(\check{g}^{cd} \partial_c a \partial_d a)}$ in \eqref{eq:Z_factor}.
The matrix $\check{g}^{ab}$ is the inverse of the following effective induced metric:
\begin{align}
\check{g}_{ab} = \Pi^{[2]}_{\mu \nu} (k_1,k_2) \partial_a X^{\mu} \partial_b
 X^{\nu} 
+ \frac{\lambda^2}{\det h^{[2]}} g_{\mu \nu} 
(k_1^{\mu} \partial_a \varphi_2 - k_2^{\mu} \partial_a \varphi_1) 
(k_1^{\nu} \partial_b \varphi_2 - k_2^{\nu} \partial_b \varphi_1).
\end{align}
It is convenient to consider the following quantity rather than the
induced metric itself:
\begin{align}
\check{g}_{ab} +
 \frac{\lambda^2 e^{2\phi}}{\det h^{[2]}} \check{F}_a^{(1)} \check{F}_b^{(1)}
\, .
\label{eq:IIA_metric}
\end{align}
The inverse of the above matrix is evaluated as 
\begin{align}
\check{g}^{ab} - 
\frac{\lambda^2 e^{2\phi} \check{F}^{(1)}_c \check{F}^{(1)}_d
 \check{g}^{ac} \check{g}^{bd} }{ (\det h^{[2]}) + \lambda^2 e^{2\phi}
 \check{F}^{(1)}_e \check{F}^{(1)}_f \check{g}^{ef}}
\, .
\label{eq:IIA_inverse_metric}
\end{align}
We have already observed that the eleven-dimensional origin of 
\eqref{eq:IIA_metric} is \eqref{eq:11d_metric} except for the dilaton factor.
Therefore the eleven-dimensional origin of \eqref{eq:IIA_inverse_metric}
is the inverse of $\hat{G}_{ab}$, namely, 
we see that $\check{g}^{ab}$ comes from $\hat{G}^{ab}$.
With these observations, we expect that the eleven-dimensional origin of $\check{g}^{ab}
\partial_a a \partial_b a$ and $\check{\mathcal{N}}$ in
\eqref{eq:Z_factor} is given by $\hat{G}^{ab} \partial_a a \partial_b a$.
Indeed, the dimensional reduction of this term gives 
\begin{align}
\hat{G}^{ab} \partial_a a \partial_b a 
=& \ e^{\frac{2}{3} \phi} (\check{g}^{ab} \partial_a a \partial_b a) 
\left[
1 - \frac{
\lambda^2 e^{2\phi} (\check{g}^{cd} \partial_c a \check{F}^{(1)}_d)^2
}{
(\check{g}^{ef} \partial_e a \partial_f a)
(\det h^{[2]} + \lambda^2 e^{2\phi} \check{g}^{gh} \check{F}^{(1)}_g \check{F}^{(1)}_h)
}
\right]
\notag \\
=& \ e^{\frac{2}{3} \phi} (\check{g}^{ab} \partial_a a \partial_b a)
 \check{\mathcal{N}}^2.
\label{eq:check_N}
\end{align}
Note that the dilaton factor $e^{\frac{2}{3} \phi}$ which
appears in \eqref{eq:check_N}, together with the factor $e^{\frac{4}{3} \phi}$ in
\eqref{eq:hat_Z_reduction}, reproduces the correct factor $e^{\phi}$ in \eqref{eq:Z_factor}.
Then we conclude that the eleven-dimensional origin of \eqref{eq:Z_factor} is 
\begin{align}
\delta_a {}^c + \frac{i}{3! \sqrt{\det \hat{h}^{[2]}}}
 \frac{1}{\sqrt{\hat{G}^{gh} \partial_g a \partial_h a}}
 \frac{\varepsilon^{gbcdef} \hat{G}_{ab} \hat{\mathcal{H}}_{def} \partial_g a}{\sqrt{ -
 \det \hat{G}}}.
\end{align}

Finally, we look for the eleven-dimensional origin of the third line in
\eqref{eq:IIA522}.
Using the induced metric $\hat{G}_{ab}$, it is straightforward to show that the following term
\begin{align}
- \frac{1}{4} T_{\text{M}5} \int \! d^6 \xi \ \frac{1}{3!}
 \frac{1}{\check{G}^{gh} \partial_g a \partial_h a} \varepsilon^{abcjkl}
 \hat{\mathcal{H}}_{jkl} \hat{\mathcal{H}}_{bcd} \partial_a a \partial_e a \, \hat{G}^{de}
\end{align}
reproduces the third line in \eqref{eq:IIA522}.

\subsubsection*{Effective action of M$5^3$-brane}
Collecting everything altogether, we conclude that the effective action of the exotic M$5^3$-brane in the
eleven-dimensional background $\hat{g}_{MN}$ is obtained as 
\begin{align}
S_{5^3} =& \ - T_{\text{M}5} \int \! d^6 \xi \ (\det \hat{h}^{[3]}) 
\sqrt{
- \det \hat{G}
}
\sqrt{
\det 
\Big(
\delta_a {}^c + \frac{i}{3! \sqrt{\det \hat{h}^{[3]}}}
 \frac{1}{\sqrt{\hat{G}^{gh} \partial_g a \partial_h a}}
 \frac{\varepsilon^{gbcdef} \hat{G}_{ab} \hat{\mathcal{H}}_{def} \partial_g a}{\sqrt{ -
 \det \hat{G}}}
\Big)
}
\notag \\
& \ - \frac{1}{4} T_{\text{M}5} \int \! d^6 \xi \ 
\frac{\varepsilon^{abcjkl} \hat{G}^{de} \hat{\mathcal{H}}_{jkl} \hat{\mathcal{H}}_{bcd} \partial_a a
 \partial_e a}{3! \hat{G}^{gh} \partial_g a \partial_h a}.
\label{eq:53}
\end{align}
Here the induced metric $\hat{G}_{ab}$ is defined by 
\begin{align}
\hat{G}_{ab} = 
\hat{\Pi}^{[3]}_{MN} (k) \partial_a X^M \partial_b X^N + 
\frac{1}{\det \hat{h}^{[3]}} \hat{g}_{MN} (\hat{k}^M_{\hat{I}} \partial_a
 \hat{\varphi}^{\hat{I}}) (\hat{k}^N_{\hat{J}} \partial_b \hat{\varphi}^{\hat{J}}).
\end{align}
The action contains all the fields in the $\mathcal{N} = (2,0)$ tensor
multiplet and the auxiliary field and couplings to the three Killing vectors
$\hat{k}^M_{\hat{I}}$ associated with the $U(1)^3$ isometry.
In an appropriate coordinate system, it is obvious that the $\det
\hat{h}^{[3]}$ gives the volume of the 3-torus $T^3$ in which the
$U(1)^3$ isometry is realized.
Therefore, the effective tension of the M$5^3$-brane is correctly
reproduced in the action. We have also confirmed that the action
\eqref{eq:53} precisely reproduces the type IIA $5^2_2$-brane action
\eqref{eq:IIA522} after the direct dimensional reduction.

Since the M5- and the M$5^3$-branes solutions are related by the
U-duality \cite{LozanoTellechea:2000mc}, 
it is worthwhile to examine the relation between the action of the M$5^3$-brane
\eqref{eq:53} and that of the M5-brane \eqref{eq:M5_action}.
If we redefine the induced metric $\hat{G}_{ab}$ as 
\begin{align}
\hat{G}^{\prime}_{ab} = (\det \hat{h}^{[3]})^{\frac{1}{3}} \hat{G}_{ab},
\end{align}
the effective action \eqref{eq:53} is rewritten as 
\begin{align}
S_{5^3} =& \  - T_{\text{M}5} \int \! d^6 \xi \ 
\sqrt{
- \det 
\Big(
\hat{G}'_{ab} + \frac{i}{3!} \frac{1}{\sqrt{-\hat{G}'}} 
\frac{1}{\sqrt{\hat{G}^{\prime gh} \partial_g a \partial_h a}} 
\varepsilon_{ab} {}^{cdef} \hat{\mathcal{H}}_{def} \partial_c a
\Big)
}
\notag \\
& \ 
- \frac{1}{4} T_{\text{M}5} \int \! d^6 \xi \ 
\frac{\varepsilon^{abcjkl} \hat{G}^{\prime de} \hat{\mathcal{H}}_{jkl} \hat{\mathcal{H}}_{bcd} \partial_a a
 \partial_e a}{3! \hat{G}^{\prime gh} \partial_g a \partial_h a}.
\label{eq:53r}
\end{align}
Here all the indices in the world-volume are raised and lowered by
$\hat{G}'_{ab}$ and its inverse $\hat{G}^{\prime ab}$.
The action \eqref{eq:53r} is nothing but the PST action \eqref{eq:M5_action} for the M5-brane 
in which the induced metric $\hat{g}_{ab}$ and $\hat{H}_{abc}$ are
replaced by $\hat{G}'_{ab}$ and $\hat{\mathcal{H}}_{abc}$ respectively.
Exploiting this fact, we find the following effective T-duality rule in
eleven dimensions:
\begin{align}
P[\hat{g}_{MN}]_{ab} 
\xrightarrow[k_1, k_2, k_3]{\text{T}} & \ 
(\det \hat{h}^{[3]})^{\frac{1}{3}} 
\left(
\hat{\Pi}^{[3]}_{MN} (k) \partial_a X^M \partial_b X^N + 
\frac{1}{\det \hat{h}^{[3]}} \hat{g}_{MN} (\hat{k}^M_{\hat{I}} \partial_a
 \hat{\varphi}^{\hat{I}}) (\hat{k}^N_{\hat{J}} \partial_b \hat{\varphi}^{\hat{J}})
\right), 
\notag \\
\hat{H}_{abc} \xrightarrow[k_1, k_2, k_3]{\text{T}} & \ \hat{\mathcal{H}}_{abc},
\label{eq:T-dual}
\end{align}
where $\hat{\mathcal{H}}_{abc}$ is the 3-form in \eqref{eq:H_T-dual}.
With this result in hand, it is obvious that the symmetries of the PST action \cite{Pasti:1997gx}
is carried over to the M$5^3$-brane action \eqref{eq:53}.

We note that the T-duality transformations \eqref{eq:T-dual} are
obtained by first compactifying along the M-circle direction, then
performing the repeated T-duality transformations in type IIA theory,
then uplifting to eleven dimensions.
Therefore the formula \eqref{eq:T-dual} should be deduced from the
$SL(3, \mathbb{Z}) \times SL(2, \mathbb{Z})$ U-duality transformations
in the based eight-dimensional theory.

\section{Source terms of the exotic five-branes}
In this section, we investigate the equations of motion for supergravity
theories in the presence of the exotic branes.
We will show that the world-volume effective action \eqref{eq:53}
provides the delta function source of the geometry near the core of
the brane. The analysis presented here is analogous to that for the
KK6-brane \cite{Bergshoeff:1997gy}.
Before we discuss the exotic M$5^3$-brane in eleven dimensions, let us 
begin with the $5^2_2$-brane solution in ten-dimensional supergravity.

\subsection{Source terms of the exotic $5^2_2$-brane}
Since the exotic $5^2_2$-brane is a solution to the NS-NS sector of 
type II supergravities \cite{LozanoTellechea:2000mc,deBoer:2010ud,deBoer:2012ma}, we consider the 
following bulk supergravity action in the Einstein frame:
\begin{align}
S^{10}_{\text{bulk}}=
{1\over \kappa^2}\int  d^{10}x \sqrt{-g^{\text{E}}}
\Big( 
R-{1\over2}\del_{\mu} \phi \del^{\mu} \phi
- \frac{1}{12}e^{-\phi} H_{\mu \nu \rho} H^{\mu \nu \rho}
\Big)
\, , \label{10D-bulk}
\end{align}
where 
$\kappa$ is the gravitational constant in ten dimensions and
$g^{\text{E}}_{\mu \nu}$ is the metric in the Einstein frame which is
defined through the metric in the string frame by $g^{\text{E}}_{\mu \nu} =
e^{- \phi/2} g_{\mu \nu}$.
$R$ is the Ricci scalar constructed by $g^{\text{E}}_{\mu \nu}$, and
$H_{\mu \nu \rho}$ is the field strength of the B-field.
To make contact with the exotic $5^2_2$-brane solution, we consider the following ansatz:
\bsubeq \label{522-BG}
\begin{align}
 d s^2_{5^2_2}
&=
 d x^2_{012345}
+ {\cal H} d x^2_{67}
+\frac{\cal H}{\cal K}
\big\{ (d x^8)^2 + (d x^9)^2 \big\}
\, , \\
{\cal K}&={\cal H}^2+\omega^2, ~~~~~e^{2\phi}={\cal H\over K},~~~~~
B_{MN}=\left\{
{\renewcommand{\arraystretch}{1.4}
\begin{array}{l}
\dps B_{89}=-{\omega \over {\cal K}}, \\
\dps 0~~~~~
\text{(otherwise)} \, ,
\end{array}
}
\right. 
\end{align}
\esubeq
where, the $5^2_2$-brane world-volume is extended along the
012345-directions.
The functions ${\cal H}$ and $\omega$ depend only on the transverse directions
$x^6$ and $x^7$:
\begin{align}
{\cal H}={\cal H}(x^6,x^7)
\, , \ls
\omega = \omega (x^6,x^7)
\, . 
\end{align}
The $U(1)^2$ isometry is taken along the 89-directions.
The derivative of ${\cal H}$ is related to the derivative of $\omega$ in the following way:
\begin{align}
{\cal H}_{,6}=\omega_{,7},~~{\cal H}_{,7}=-\omega_{,6}, \label{Homega}
\end{align}
where ${\cal H}_{,m}$ implies the derivative of ${\cal H}$ with respect
to $x^m$, i.e., ${\cal H}_{,m} = \del_m {\cal H}$.
This relation comes from the Dirac monopole equation.
We will find that ${\cal H}$ is a harmonic function on the 67-plane by solving the equation
of motion for each field.

The variations of the bulk action (\ref{10D-bulk}) with respect to
the dilaton and the metric $g^{\text{E}}_{\mu \nu}$ together with the
ansatz \eqref{522-BG} are evaluated to be
\bsubeq
\begin{align}
\delta S^{10}_{\text{bulk}} \Big|_{\phi}
&=
{1\over\kappa^2} \int d^{10}x \sqrt{-g^{\text{E}}} \,
\Bigg\{
\frac{1}{2 {\cal K}^2} \bigg({\cal K\over H} \bigg)^{7\over4}
\biggl(1- \frac{2 {\cal H}^2}{{\cal K}}\biggr)({\cal H}_{,66}+{\cal H}_{,77})
\Bigg\}
\, \delta \phi 
\, , \label{10dimansionaleqdilaton}\\
\delta S^{10}_{\text{bulk}}|_{g^{\text{E}}_{\mu \nu}}
&=
-{1\over\kappa^2} \int d^{10}x \sqrt{-g^{\text{E}}} \,
G^{\mu \nu} 
\, \delta g^{\text{E}}_{\mu \nu} 
\, , \label{EQuationformetric}\\
G^{\mu \nu}
&=
\left\{
{\renewcommand{\arraystretch}{2.0}
\begin{array}{rcl}
\dps 
G^{ab}
\!\!&=&\!\!
\dps {\eta^{ab}\over2{\cal H}^2{\cal K}^2}\sqrt{{\cal H}\over {\cal K}}
\Big\{ 
-2{\cal K}({\cal H}_{,6}{\cal K}_{,6}+{\cal H}_{,7}{\cal K}_{,7})+
{\cal H}({\cal K}^2_{,6}+{\cal K}_{,7}^2)+{\cal K}^2({\cal H}_{,66}+{\cal H}_{,77})
\Big\}
\, , \\
\dps 
G^{IJ}
\!\!&=&\!\!
\dps \frac{\delta^{IJ}}{2 {\cal K}^3}\biggl({{\cal K}\over {\cal H}}\biggr)^{3\over2}
\Big\{ -{\cal K}^2_{,6}-{\cal K}^2_{,7}+{\cal K}({\cal K}_{,66}+{\cal K}_{,77})\Big\}
\, ,\\
\multicolumn{3}{l}{0~~~~~\text{(otherwise)} \, ,}
\end{array}
}
\right.
\end{align}
\esubeq
where $G^{\mu \nu}$ is the Einstein tensor
whose indices are decomposed into three parts, i.e.,
\begin{align}
\mu , \nu \ = \ 0,1,\ldots, 9
\, , \ls
a, b \ = \ 0, 1, \ldots, 5
\, , \ls
m,n \ = \ 6,7
\, , \ls
I, J \ = \ 8,9
\, . 
\end{align}
We notice that the index $I$ runs 1 and 2 in the previous section, which merely implies the counting the number of the Killing vectors.
Here $I$ denotes the 8- or 9-th direction in the ten-dimensional spacetime. 
We emphasize that the equation of motion for the B-field in the bulk
supergravity action (\ref{10D-bulk}) is trivially satisfied due to the relation (\ref{Homega}).

We now introduce the type IIA $5^2_2$-brane action \eqref{eq:IIA522} to the bulk
action \eqref{10D-bulk}. The total action we consider is
\begin{align}
S = S^{10}_{\text{bulk}} + S^{\text{IIA}}_{5^2_2}.
\label{eq:action}
\end{align}
Since we are interested in the NS-NS sector, we 
remove all the R-R fields in the action \eqref{eq:IIA522}. 
We also ignore the fluctuations of the 
world-volume fields $Y$, $\varphi_{I}$ and $A_{ab}$ because they are small perturbations to the
geometry. Then, we find $\check{H}_{abc} = 0$.
Therefore, only the non-trivial part of the action \eqref{eq:IIA522} is
\begin{align}
S_{5^2_2}^{\text{IIA}} = - T_5 \int \! d^6 \xi \ (\det h^{[2]}) e^{-2
 \phi} 
\sqrt{- \det \check{g}_{ab}
},
\label{eq:IIA522_2}
\end{align}
where the T-dualized metric is evaluated as
\begin{align}
\check{g}_{ab} = g_{ab} - \frac{1}{2} h^{[2]IJ} 
\Big\{ (i_{k_I} g - i_{k_I} B)_a (i_{k_J} g +
 i_{k_J} B)_b + (a \leftrightarrow b) \Big\} .
\end{align}
From now on, we employ the static gauge where the world-volume
coordinates $\xi^a$ are identified with the space-time coordinates
$X^a$.
The pull-back of a bulk field $\Phi_{\mu \nu}$ in the action
\eqref{eq:IIA522_2} is evaluated as
\begin{align}
P[\Phi]_{ab} = \Phi_{ab} + \Phi_{am} \partial_b X^m 
+ \Phi_{mb} \partial_a X^m + \Phi_{mn} \partial_a X^{m} \partial_b
 X^{n}.
\end{align}
We note that the fluctuation modes $X^I$ along the isometry directions
have been identified with the winding modes $\varphi_1$, $\varphi_2$ 
which are removed from the action \eqref{eq:IIA522}.
We now consider the configuration $X^m = 0$ and take the Killing vectors as
\begin{align}
k^{\mu}_1 = \delta^{\mu}_8, \quad k^{\mu}_2 = \delta^{\mu}_9.
\end{align}
Then, using the ansatz \eqref{522-BG} for the $5^2_2$-brane background, 
the pull-back of $i_{k_I} B$ vanishes and 
we are left with the pull-back of the following quantity:
\begin{align}
\check{g}_{\mu \nu} =
 g_{\mu \nu} - h^{[2]IJ} k^{\rho}_I k^{\sigma}_J
 g_{\mu \rho} g_{\nu \sigma}
= \Pi_{\mu\nu}^{[2]} (k_1,k_2),
\end{align}
where $h^{[2]}_{IJ}$ is the Killing matrix including the B-field.

Consequently, the variations of the action \eqref{eq:IIA522_2} with
respect to the dilaton, the B-field and the metric $g_{\mu
\nu}^{\text{E}}$ are calculated to be
\bsubeq
\begin{align}
\delta S^{\text{IIA}}_{5^2_2}|_{\phi}
&=-T_{5}\int d^6 \xi~ \, 
\Big\{
-{1\over2}(\det h^{[2]})+\bigg( \frac{\cal H}{\cal K} \bigg)^2
\Big\}
\, e^{-2\phi} \, \delta\phi
, \label{sugra dilaton}\\
\delta S^{\text{IIA}}_{5^2_2}|_{g_{\mu \nu}^{\text{E}}}
&=
-T_{5}\int d^6 \xi \, 
e^{-{3\over2}\phi} (\det h^{[2]}) 
    \Big\{ k^{\mu}_I k^{\nu}_J(h^{[2]JI} +{1\over2}\eta^{ab}\delta^{\mu}_a \delta^{\nu}_b) \Big\}
\delta g^{\text{E}}_{\mu \nu} 
, \label{522grav}\\
\delta S^{\text{IIA}}_{5^2_2}|_{B_{\mu \nu}}
&=
-T_{5}\int d^6 \xi \, 
\Big\{ e^{-2\phi}(\det h^{[2]}) h^{[2]JI} k^{\mu}_I k^{\nu}_J
\Big\} \delta B_{\mu \nu}.
 \label{sugra Bba}
\end{align}
\esubeq

Now, we evaluate the equations of motion derived from the total action \eqref{eq:action}.
The equation of motion for the dilaton is obtained from (\ref{10dimansionaleqdilaton})
and (\ref{sugra dilaton}):
\begin{align}
{1\over2 {\cal K}^2}\biggl({\cal K\over H}\biggr)^{7\over4}
\biggl(1-{2 {\cal H}^2\over{\cal K}}\biggr)({\cal H}_{,66}+{\cal H}_{,77})
=
{\kappa^2 \over \sqrt{-g^{\text{E}}}}T_{5}
~\delta^2(x) e^{-2\phi}
\Biggl[
-{1\over2}(\det h^{[2]})+\biggl({\cal H\over K} \biggr)^2
\Biggr]. \label{dilatoneQuation}
\end{align}
Here the delta function source in the right hand side comes from the second term in \eqref{eq:action}.
Taking into account the static gauge, the determinant factor $\det h^{[2]}$ is
evaluated as
\begin{align}
\det h^{[2]} &={{\cal H}^2\over{\cal K}^2}+B_{89}^2={1\over{\cal K}}. 
\end{align}
Inserting the result to (\ref{dilatoneQuation}), the equation becomes
\begin{align}
({\cal H}_{,66}+{\cal H}_{,77})&=-\kappa^2 T_{5}\delta^2(x). 
\label{DifferencialEQ}
\end{align}
The equation \eqref{DifferencialEQ} 
is nothing but the Poisson equation in two dimensions.
This equation is solved as
\begin{align}
{\cal H}={\cal H}_0+{\kappa^2 T_{5}\over2\pi} \log {\mu\over\rho},
~~~\rho^2=(x^6)^2+(x^7)^2, \label{harmonixH}
\end{align}
where ${\cal H}_0$ and $\mu$ are the integration constants.
This is the two-dimensional harmonic function. 
Since there is the relation (\ref{Homega}), the function $\omega$ is given by
\begin{align}
\omega
=
- {\kappa^2 T_{5} \over2\pi} \tan^{-1} \Big({x^7\over x^6} \Big)
. \label{omegaON}
\end{align}
The functions (\ref{harmonixH}) and (\ref{omegaON}) correctly
reproduce the known exotic $5^2_2$-brane solution \cite{deBoer:2010ud,deBoer:2012ma}.
We stress that the world-volume effective action \eqref{eq:IIA522} 
provides the delta function source in the equation
\eqref{DifferencialEQ} and the solution \eqref{522-BG} is valid deep
inside the core of the brane geometry.
The equations of motion for the metric (\ref{EQuationformetric}) and (\ref{522grav}) give the same result. 
We also note that the equation of motion for $X^m$ is automatically satisfied by the configuration $X^m = 0$.

For the equation of motion for the B-field, 
the contribution from the bulk action \eqref{10D-bulk} vanishes
but that from the world-volume action \eqref{sugra Bba} is not zero.
The resolution to this puzzle lies in the logarithmic
behaviour of the harmonic function \eqref{harmonixH}.
Since the harmonic function $\mathcal{H}$ is a part of the metric
components, this should be a positive definite function.
This implies that the solution \eqref{522-BG} is well-defined only for
the region $\rho \ll \mu$, namely, near the core of the brane.
We can see the consistency of this picture as follows. 
Let us consider the (8,9) component of the equation (\ref{sugra Bba}):
\begin{align}
(\det h^{[2]}) h^{[2]JI} k^{8}_I k^{9}_J =B_{89}=-{\omega\over{\cal K}}.
\end{align}
When we take the near brane limit $\rho\rightarrow0$, this term becomes zero because ${\cal H}$ and $\omega$ are defined in (\ref{harmonixH}) and (\ref{omegaON}): 
\begin{align}
\lim_{\rho\rightarrow0}(B_{89})=
\lim_{\rho\rightarrow0} \Big( -{\omega\over{\cal K}} \Big) =
\lim_{\rho\rightarrow0} \Big(-{\omega\over{\cal H}^2+\omega^2 } \Big)
= 0. 
\end{align}
The other components of (\ref{sugra Bba}) also become zero by the near brane limit:
\bsubeq 
\begin{align}
\lim_{\rho\rightarrow0} \Big( {\cal H\over K} \Big) &=
\lim_{\rho\rightarrow0} \Big({{\cal H}\over {\cal H}^2+\omega^2} \Big)
=
\lim_{\rho\rightarrow0} \Big( {1\over {\cal H}} \Big)
= 0
, \\
\lim_{\rho\rightarrow0}(B_{89}) &=
\lim_{\rho\rightarrow0} \Big( -{\omega\over{\cal K}} \Big)=
\lim_{\rho\rightarrow0}\Big( -{\omega\over{\cal H}^2+\omega^2} \Big)
= 0. 
\end{align}
\esubeq
From these facts, we conclude that the equation of motion for the B-field is satisfied only near the brane. 
This is consistent with the fact that the exotic brane solutions are
valid only near the brane \cite{deBoer:2010ud,deBoer:2012ma}.
We can show that the same conclusion holds even for the type IIB
$5^2_2$-brane action \cite{Chatzistavrakidis:2013jqa,Kimura:2014upa}.

\subsection{Source terms of the exotic M$5^3$-brane}

Finally, we 
study the world-volume action of the M$5^3$-brane (\ref{eq:53})
as the source of the 
background geometry of the M$5^3$-brane in M-theory. 
The bulk supergravity action in eleven dimensions is given by 
\begin{align}
S^{11}_{\text{bulk}}=
{1\over
\kappa_{11}^2}\int d^{11}x \sqrt{-
\hat{g}} \Bigl(\hat{R}-{1\over48} \hat{F}^{MNPQ} \hat{F}_{MNPQ}\Bigr)
\, ,
\label{11D-bulk}
\end{align}
where $\kappa_{11}$ is the gravitational constant in eleven dimensions,
$\hat{g}$ is the determinant of the eleven-dimensional metric
$\hat{g}_{MN}$, $\hat{R}$ is the Ricci scalar defined by $\hat{g}_{MN}$,
and $\hat{F}_{MNPQ}$ is the field strength of the 3-form field $\hat{C}^{(3)}$.
We first take a look at the structure of the equations of motion derived from
the bulk action \eqref{11D-bulk}.
We consider the following ansatz for the M$5^3$-brane solution 
\cite{LozanoTellechea:2000mc, deBoer:2010ud,deBoer:2012ma}:
\bsubeq \label{M53-BG}
\begin{align}
d s^2_{11}&=
{\cal A}
\Big\{ d x^2_{012345}+{\cal H}\bigl((d x^6)^2+(d x^7)^2\bigr)
   +{\cal H \over K} d x^2_{89\sharp}
   \Big\}
\, , \label{Sukekiyo}\\
{\cal A}&=\Bigl({\cal K\over H}\Bigr)^{1\over3}, ~~~~~ {\cal K}={\cal H}^2+\omega^2,~~~~~
\hat{F}_{MNPQ}=+\ve_{MNPQX}\partial^X\Bigl({\cal H\over K}\Bigr)
\, , \label{saruzou}
\end{align}
\esubeq
where $\ve_{MNPQX}$ is the Levi-Civita symbol whose normalization is $\ve_{6789\sharp}=+1$. 
Again the functions ${\cal H}$ and $\omega$ depend only on the coordinates $(x^6,x^7)$, and they are subject to the relation (\ref{Homega}).
Using the ansatz (\ref{M53-BG}), the variation of the action \eqref{11D-bulk} with respect to the metric is
\bsubeq
\begin{align}
\delta S^{11}_{\text{bulk}} \Big|_{g_{MN}}&=
-{1\over\kappa_{11}^2}\int d^{11}x \sqrt{-\hat{g}} \, 
G^{MN} \, \delta \hat{g}_{MN}
\, , \label{11Dgrav}\\
G^{MN}&=
\left\{
{\renewcommand{\arraystretch}{2.0}
\begin{array}{rcl}
G^{ab}
\!\!&=&\!\!
\dps 
\frac{\eta^{ab}}{2{\cal H}^2}
\Big( \frac{\cal H}{\cal K} \Big)^{2\over3}
({\cal H}_{,66}+{\cal H}_{,77})
\, , \\
G^{\hat{I} \hat{J}}
\!\!&=&\!\!
\dps
\frac{\delta^{\hat{I} \hat{J}}}{\cal H}
\Big( \frac{\cal H}{\cal K} \Big)^{2\over3}({\cal H}_{,66}+{\cal H}_{,77})
\, ,\\
\multicolumn{3}{l}{0~~~~~\text{(otherwise)}\, ,}
\end{array}
}
\right.
\end{align}
\esubeq
where $G^{MN}$ is the Einstein tensor in eleven dimensions.
Here we again split the eleven-dimensional indices as follows:
\begin{align}
M, N \ = \ 0,1,\ldots, 9, \sharp
\, , \ls
a,b \ = \ 0,1,\ldots, 5
\, , \ls
m,n \ = \ 6,7
\, , \ls
\hat{I}, \hat{J} \ = \ 8,9, \sharp
\, .
\end{align}
The equation of motion for the 3-form field $\hat{C}^{(3)}_{MNP}$
is trivial under the ansatz (\ref{saruzou}):
\begin{align}
\delta S^{11}_{\text{bulk}} \Big|_{
C^{(3)}_{MNP}
}&=0 \, . 
\end{align}

We next introduce
the M$5^3$-brane world-volume action obtained in (\ref{eq:53}) and consider the total action
\begin{align}
S = S^{11}_{\text{bulk}} + S_{5^3}.
\label{eq:11D_total_action}
\end{align}
We ignore the scalar fields
$\hat{\varphi}_{1,2,3}$, and the 2-form field $A_{ab}$, 
then the action \eqref{eq:53} is reduced to
\bsubeq
\begin{align}
S_{5^3}&=-T_{\text{M5}}\int d^6\xi(\det \hat{h}^{[3]})\sqrt{-\det \hat{G}_{ab}} 
\, , \label{53action}\\
\hat{G}_{ab}&=\hat{\Pi}_{MN}^{[3]} (k) \partial_aX^M\partial_bX^N 
\, . \label{53-sayou}
\end{align}
\esubeq
Since the action contains only the metric $\hat{g}_{MN}$ and the 
geometric zero-modes $X^M$ on the brane,
the action has no contribution to the field equation for the 3-form
field $\hat{C}^{(3)}$.

Following the analysis made in the $5^2_2$-brane case, we employ the
static gauge and remove the geometric zero-modes
\begin{align}
\xi^a=X^a,
~~~X^m=0.
\label{static-M}
\end{align}
We take the Killing vectors as
\begin{align}
\hat{k}^M_1 = \delta^M_8, \quad
\hat{k}^M_2 = \delta^M_9, \quad
\hat{k}^M_3 = \delta^M_{\sharp}.
\end{align}
Furthermore, using the ansatz (\ref{M53-BG}),
$\hat{h}^{[3]}_{\hat{I} \hat{J}}$,  $\hat{G}_{ab}$ and $\sqrt{-\det \hat{G}}$ are explicitly given as 
\begin{align}
\hat{h}^{[3]}_{\hat{I} \hat{J}} 
={\cal A}{\cal H\over K}\delta_{IJ}, ~~~~~
\hat{G}_{ab}
=
\hat{g}_{ab}
~~~~~
\sqrt{-\det \hat{G}} 
={\cal K \over H}
\, .
\end{align}
By virtue of the static gauge (\ref{static-M}) and the above reduction,
we derive the equations of motion for the metric and the geometric zero-modes $X^M$
from the world-volume action (\ref{53action}). 
Since the equation of motion for $X^M$ becomes trivial under the
condition \eqref{static-M},
it is enough to check the equation of motion for the metric.
The variation of the action \eqref{53action} with respect to the metric
$\hat{g}_{MN}$ together with the ansatz (\ref{M53-BG})
is calculated to be
\begin{align}
\delta S_{5^3}&=-T_{\text{M5}}\int d^6 \xi 
\Big\{
\Big( \frac{\cal H}{\cal K} \Big)^{1\over3} \, \hat{k}^M_{\hat{I}} \hat{k}^N_{\hat{J}} \delta^{\hat{I} \hat{J}}
+{1\over2}\Bigl({\cal H\over K}\Bigr)^{4\over3} \eta^{ab}\delta^{M}_a\delta^{N}_b
\Big\}
\, \delta 
\hat{g}_{MN}
\, . \label{53branegrav}
\end{align}
Combining the results (\ref{11Dgrav}) and (\ref{53branegrav}), 
we obtain the equation of motion derived from the total action
\eqref{eq:11D_total_action} as
\begin{align}
G^{MN}
=
-
{\kappa_{11}^2 T_{\text{M5}}\over \sqrt{-\hat{g}}}
\Big\{
\Big( \frac{\cal H}{\cal K} \Big)^{1\over3} \hat{k}^{M}_{\hat{I}} \hat{k}^{N}_{\hat{J}} \delta^{\hat{I} \hat{J}}
+{1\over2}\Bigl({\cal H\over K}\Bigr)^{4\over3} \eta^{ab}\delta^{M}_a \delta^{N}_b
\Big\} \delta^2(x) 
\, . \label{Finalequation}
\end{align}
The non-zero quantities of the above
equation are divided into two parts. 
First, we consider the case for $(M,N)=(\hat{I}, \hat{J})$. 
In this case, the left- and the right-hand 
sides of the equation (\ref{Finalequation}) are calculated as 
\bsubeq
\begin{align}
\mbox{(LHS)}&={1\over{\cal H}}\Bigl({\cal H \over K}\Bigr)^{2\over3}({\cal H}_{,66}+{\cal H}_{,77})\delta^{\hat{I} \hat{J}}
\, , \\
\mbox{(RHS)}&= 
-\frac{\kappa_{11}^2 T_{\text{M5}}}{\cal H} \Bigl({\cal H \over K}\Bigr)^{2\over3} \delta^2(x)\delta^{\hat{I} \hat{J}}
\, . 
\end{align}
\esubeq
These two equations imply that the function $\mathcal{H}$ satisfies the following equation:
\begin{align}
({\cal H}_{,66}+{\cal H}_{,77})=
-\kappa_{11}^2 T_{\text{M5}}\delta^2(x). \label{53Harmonic}
\end{align}
The equation (\ref{53Harmonic}) is the same as the differential equation which we found in (\ref{DifferencialEQ}), 
and the function ${\cal H}$ is solved as
\begin{align}
{\cal H}={\cal H}_0+{\kappa_{11}^2 T_{\text{M5}}\over 2\pi}\log
{\mu\over \rho},
~~~\rho^2=(x^6)^2+(x^7)^2. 
\label{eq:53harmonic}
\end{align}
The result correctly reproduces the known solution of the M$5^3$-brane.
The delta function in \eqref{53Harmonic}, which comes from the second term in
\eqref{eq:11D_total_action}, allow the solution to be defined at the
core of the brane geometry.
Next, we consider the case for $(M,N)=(a,b)$. 
The left- and the right-hand sides of  (\ref{Finalequation}) are 
\bsubeq
\begin{align}
\mbox{(LHS)}&={1\over2{\cal H}^2}\Bigl({\cal H \over K}\Bigr)^{2\over3}({\cal H}_{,66}+{\cal H}_{,77})\eta^{ab}
\, , \\
\mbox{(RHS)}&=
-\frac{\kappa_{11}^2 T_{5^3}}{2 {\cal K}} \Bigl({\cal H\over K}\Bigr)^{2\over3}\eta^{ab}\delta^2(x) 
\, . 
\end{align}
\esubeq
Then we find the following equation:
\begin{align}
({\cal H}_{,66}+{\cal H}_{,77})=
- \kappa_{11}^2 T_{\text{M5}}
\frac{{\cal H}^2}{\cal K}
\delta^2(x). \label{ijindexes}
\end{align}
This equation is slightly different from (\ref{53Harmonic}).
At first sight, the equations (\ref{53Harmonic}) and (\ref{ijindexes})
look inconsistent. However, we have encountered the same situation in
the type IIA $5^2_2$-brane case. We should take care of the 
validity of the solution.
The logarithmic behaviour of the harmonic function \eqref{eq:53harmonic}
indicates that the solution becomes the most rigorous at the core of the
brane geometry. Indeed, when we take the near brane limit
$\rho\rightarrow0$, then the equation (\ref{ijindexes}) is drastically simplified:
\begin{align}
\lim_{\rho\rightarrow0}{{\cal H}^2\over {\cal K}}=
\lim_{\rho\rightarrow0}{{\cal H}^2\over {\cal H}^2+\omega^2}
=
{{\cal H}^2\over {\cal H}^2}=1.
\end{align}
Thus, we conclude that (\ref{ijindexes}) is equal to (\ref{53Harmonic}) with the near brane limit and the function ${\cal H}$ again reproduces the M$5^3$-brane solution.

From these results, we conclude that the world-volume action of the M$5^3$-brane is the source of the M$5^3$-brane.

\section{Conclusion and discussions}

In this paper we studied the world-volume effective action of the
exotic five-brane, which we call the M$5^3$-brane, in eleven dimensions.
The background geometry of the M$5^3$-brane is an U-fold in eleven
dimensions which has three isometry directions in the transverse space
to the world-volume.
The world-volume fields in the action consists of the six-dimensional
$\mathcal{N} = (2,0)$ tensor multiplet, including two geometric
zero-modes, three scalar fields, a self-dual field and their
superpartners. The action contains three Killing vectors
$k^{M}_{\hat{I}} \ (\hat{I} = 1,2,3)$ associated with the $U(1)^3$ isometry.
We showed that the direct dimensional reduction of the M$5^3$-brane action reproduces
that of the exotic $5^2_2$-brane in type IIA string theory.
We found that the world-volume effective action of the 
M$5^3$-brane is obtained from that of the M5-brane by the T-duality transformation of
the background fields.
We explicitly wrote down the effective T-duality rule in eleven dimensions.
The transformation is generated by a subgroup of the $SL(3,\mathbb{Z}) \times SL(2,\mathbb{Z})$ U-duality group.

We also demonstrated that our action \eqref{eq:53} 
provides the source term of the M$5^3$-brane geometry near the core of the brane. The $\det
\hat{h}^{[3]}$ factor, which is necessary for the appropriate tension, 
 plays a crucial role in the supergravity equation of motion.
Although the exotic branes are non-geometric objects 
\cite{Hull:2004in,deBoer:2010ud,deBoer:2012ma}, 
these results reflect the fact that the exotic branes are genuine
``branes'' where something of the energy lump is localized around the
center of the brane geometry.
However, we showed that the source description is available only near the core of the
brane geometry. This is conceivable since the exotic $5^2_2$-brane is not
well-defined as a stand alone object \cite{deBoer:2010ud,deBoer:2012ma}.
In order to construct the asymptotically flat solution,
we need to consider other duality branes simultaneously 
in addition to the $5^2_2$-brane.
Indeed, the solution \eqref{522-BG} is just the near center limit to the
$5^2_2$-brane of the asymptotically flat solution \cite{Kikuchi:2012za}.
The integration constant $\mu$ is the cutoff scale where the solution
fail to describe the isolated brane.
The same is true for the exotic M$5^3$-brane.
The logarithmic behavior of the harmonic function indicates that the
solution \eqref{M53-BG} is valid only near the core of the brane.

A few comments are in order.
We found the effective action of the M$5^3$-brane in the purely
geometric background, namely, the background field is only the metric.
It is important to find the coupling to the supergravity 3-form field
(C-field). 
We encountered the terms including $\mathcal{G}_{Ia}$ which come from the T-duality
transformation of the C-field.
The three Killing vectors $\hat{k}^M_{\hat{I}}$ in \eqref{eq:H_T-dual}
do not appear on an equal footing.
Since the T-duality transformation of the C-field becomes
quite non-linear, it is not straightforward to write it down in a
covariant fashion.
In this paper, we worked out only the T-duality 
subgroup of the $SL(3,\mathbb{Z}) \times SL(2,\mathbb{Z})$ U-duality group
and the Killing vectors are always in the transverse directions to the brane world-volume.
Therefore our action \eqref{eq:53} never knows about the exotic 
$4^3_3$-brane which should be obtained by the double dimensional reduction of the
M$5^3$-brane. In order to construct the complete action of the M$5^3$-brane,
one needs full of the $SL(3,\mathbb{Z}) \times SL(2, \mathbb{Z})$ U-duality transformation rule.
The exceptional field theory approach 
\cite{Hull:2007zu, Thompson:2011uw, Berman:2011jh, Berman:2013eva,
Hohm:2013pua, Blair:2013gqa, Blair:2014zba, Hohm:2015xna, Berman:2014jsa}
may be helpful in this direction.
The Wess-Zumino part of the action is also important.
Since the exotic branes are sources of the non-geometric fluxes
\cite{Hassler:2013wsa, Andriot:2014uda, Sakatani:2014hba},
the Wess-Zumino term of the M$5^3$-brane action contains couplings of the
non-geometric flux to the M$5^3$-brane.
We will come back to these issues in future studies.

\subsection*{Acknowledgments}

T.~K. is supported by the MEXT-Supported Program for the Strategic Research Foundation at Private Universities ``Topological Science'' ({Grant No.~S1511006}). 
He is also supported in part by the Iwanami-Fujukai Foundation.
The work of S.~S. is supported in part by Kitasato University Research Grant for Young
Researchers.
The work of M.~Y. is supported by NUS Tier 1 FRC Grant R-144-000-316-112.


\end{document}